\documentclass[submission]{eptcs}
\providecommand{\event}{F-IDE 2015}

\usepackage{siunitx}
\usepackage{tikz}
\usetikzlibrary{shapes.misc,shapes.geometric,positioning,arrows,fit,backgrounds}
\usepackage{listings}
\usepackage{hyperref}

\newcommand{\doi}[1]{\textsc{doi}:\href{http://dx.doi.org/#1}{\nolinkurl{#1}}}

\vfuzz \hfuzz
\lstdefinestyle{acsl-block}{
  emph=[1]{assert, assumes, assigns, axiom, axiomatic, decreases,
    ensures, ghost, invariant, lemma, logic, loop, predicate, reads,
    requires, variant}, 
  emphstyle=[1]{\bfseries},
  emph=[2]{behavior, behaviors, complete, disjoint, for:},
  emphstyle=[2]{\bfseries},
  emph=[3]{typedef, int, char, integer, real, bool, size_type, value_type,
           uint8_t, uint16_t},
  emphstyle=[3]{\bfseries},
  escapeinside={//`}{`//},
  morecomment=*[l]{@},
  morecomment=*[s]{/*@}{*/},
  moredelim=*[is][\bfseries]{|*}{*|},
}

\title{An experimental Study using ACSL and Frama-C to formulate and
  verify Low-Level Requirements from a DO-178C compliant Avionics
  Project} 
\author{Frank Dordowsky
\institute{ESG Elektroniksystem- und Logistik GmbH, 82256
  F\"urstenfeldbruck, Germany}
\email{frank.dordowsky@esg.de}}

\def\titlerunning{ACSL and Frama-C}
\def\authorrunning{F. Dordowsky}

\begin{document}
\maketitle 

\begin{abstract}
  Safety critical avionics software is a natural application area for
  formal verification. This is reflected in the formal method's
  inclusion into the certification guideline DO-178C and its formal
  methods supplement DO-333. Airbus and Dassault-Aviation, for
  example, have conducted studies in using formal verification. A
  large German national research project, Verisoft XT,
  also examined the application of formal methods in the avionics
  domain.

  However, formal methods are not yet mainstream, and it is
  questionable if formal verification, especially formal deduction,
  can be integrated into the software development processes of a
  resource constrained small or medium enterprise (SME). ESG, a Munich
  based medium sized company, has conducted a small experimental study
  on the application of formal verification on a small portion of a
  real avionics project. The low level specification of a software
  function was formalized with ACSL, and the corresponding source code
  was partially verified using Frama-C and the WP plugin, with
  Alt-Ergo as automated prover.

  We established a couple of criteria which a method should meet to be
  fit for purpose for industrial use in SME, and evaluated these
  criteria with the experience gathered by using ACSL with Frama-C on
  a real world example. The paper reports on the results of this study
  but also highlights some issues regarding the method in general
  which, in our view, will typically arise when using the method in
  the domain of embedded real-time programming.
\end{abstract}

\section{Introduction}
\label{sec:introduction}

Recent advances in automated theorem provers have brought formal
methods from purely academic exercises close to industrial use.
Airborne software has early been recognized as a suitable candidate
for the application of formal methods \cite{Rushby1993,Vries1996}.
Airbus, for example, has examined the formal method Caveat to replace
unit tests \cite{Souyris2009}. Since Airbus is a trend setter in the
avionics industry at least in Europe, one can expect the obligation to
use formal methods for the development of highly safety critical
airborne software in future. Moreover, standard organizations start
to incorporate formal methods into their regulations as one can
already see with the formal methods supplement DO-333 \cite{DO333} of
the new version of the avionics software certification guidelines,
DO-178C \cite{DO178C}. Another well-known example is the Common
Criteria for Information Technology Security Evaluation that demand
the use of formal methods for the highest Evaluation Assurance Levels
EAL~6 and EAL~7 \cite{CCPart3}. It is probably safe to assume that more
and more development standards will mandate the application of formal
methods in the future. This will force also small and medium
enterprises (SME) that develop safety critical software to consider
the adoption of formal methods in the long run.

ESG as a medium sized company that develops airborne system solutions
has therefore launched a small experimental study to examine if formal
verification can be integrated into its development processes for
airborne software, and to identify the prerequisites for an adoption
of formal methods in future DO-178C compliant software projects.

In Germany, the Verisoft XT project \cite{Baumann2009} examined the
application of formal methods in an industrial context. ESG
participated in a work package of that project that examined the
application of formal methods in the avionics domain
\cite{Blasum2012}. This experimental study is a continuation of
ESG's work in that project.

ACSL is an acronym for ``ANSI C Specification Language'' provided for
the open source tool platform Frama-C \cite{ACSLWeb,Correnson2013}.
It is a behavioral specification language that can express properties
of C source code including pre-conditions and post-conditions of C
functions using first order logic. The ACSL specifications are
provided as annotations to the source code. The WP plug-in of the
framework allows a deductive verification of the source code against
the formal annotations in ACSL \cite{Baudin2013a}.

DO-178C defines \emph{Low-Level Requirements (LLR)} as software
requirements from which source code can be directly implemented
without further information. In the software projects we
have been involved in, these requirements were natural
language annotations to the subroutines or function declarations
provided by the interfaces of the software modules. It is therefore
natural to consider ACSL as a candidate to formally express the
low-level requirements.

Several so-called DO-178C \emph{objectives} are associated with
low-level requirements that must be fulfilled for acceptance of the
software by certification authorities. One such objective is the
verifiability of low-level requirements. This objective is indeed
achieved when a formal notation is used. Another objective is the
compliance of the source code with the low-level requirements, which is
usually shown by code review. These time consuming reviews may be
replaced by automated formal verification. Blasum et al
\cite{Blasum2012} discuss these and other DO-178C objectives for the 
 formal methods of the Verisoft XT project.

The major goal of the study was to check if formal notations
can be used in a real project to formally express real world low level
requirements, i.e. to see if a certain formal method is fit for use in
industrial practice in general and for ESG in particular. A secondary
goal was to examine if available tools can verify such annotated code
and even be able to find bugs not yet discovered by testing, which would
prove one of the claimed advantages of formal methods over testing. 

The first step was to establish criteria which a formal notation and
its supporting tools should meet in order to be ready for industrial
use in an SME setting. In a second step,  we selected
a function from a real avionics project and formalized the natural
language specifications of the associated C functions into ACSL
specifications. It was in this step that we
encountered the first obstacles although the selected function was
rather trivial.

These obstacles are, in our view, quite typical for
embedded real-time programming. The problems were solved with support
of Frama-C experts via the Frama-C mailing list. However, the
solutions are not fully satisfactory which may be inherent to the
approach, as it will be discussed later.

In a next step, we attempted to verify the source code against
the ACSL specification while instrumenting the source code with
assertions in order to guide the prover. Not all verification attempts
were successful. We were only able to explain some
of the failed attempts before running out of time\footnote{Ironically,
  the failed proof attempts that we were not able to explain
  were those where the prover timed out.}. In a final step, 
we compared our experience in using tool and method against
the formerly established criteria.

\section{Industrial Fit for Purpose}
\label{sec:industr-fit-purp}

The main goal of the study was to check if formal specification and
verification is already suitable to be used in an industrial rather
than academic context, in a small to medium sized enterprise. In order
to decide on this question, a number of criteria must be fulfilled:

\begin{itemize}
\item Notation and method should be within the normal range of
  experience of an average software engineer, i.e. it should be usable
  for engineers without a PhD in formal logic.
\item The complexity of the formal language should have the same level
  as a programming language such as Ada or C. The learning curve
  should be moderate.
\item Training courses should be available.
\item Information sources such as text books, guidelines, etc. should
  be available.
\item The tools used for specification and verification should be
  mature and stable.
\item The tool should provide feedback to the user in case a proof
  fails, ideally by showing the places where the proof fails together
  with a counter example.
\end{itemize}

For avionics projects that are going to be developed in accordance
with DO-178C, additional criteria derived from DO-333 objectives such
as method soundness, coverage criteria or tool qualification have to
be considered. Such considerations were not in the scope of this
study. For a discussion of these criteria for VCC \cite{VCCWebsite}, a
tool similar to Frama-C, see \cite{Blasum2012}.

\section{Frama-C, WP Plugin and Alt-Ergo}
\label{sec:frama-c}

Avionics software is usually real-time, embedded, i.e. hardware
related, resource constrained software that often implements complex
arithmetic algorithms. One will find type casts, especially between
integer and hardware addresses, and low level pointer handling, as
well as floating point calculations.

Embedded projects often do not use an operating system so that the
developer must directly access and manipulate the hardware.

The strict real-time requirements sometimes lead to the application of
the most efficient programming constructs that may be difficult to
verify with formal methods.

A formal notation and its supporting tools must address these
constraints. This is the case with the open source tool platform
Frama-C together with its WP plugin. Frama-C is organized into a
plug-in architecture with a common kernel that allows the plug-ins to
interact with each other using ACSL as lingua franca.

The WP plugin was selected because it well supports the aforementioned
properties of avionics software. WP uses memory and arithmetic models
in order to model memory management of dynamic structures (pointers)
and properties of integer and real variables \cite{Baudin2013a}.

\section{Function Selection}
\label{sec:function-selection}

The examples in this study have been taken from the control software
of a sensor that ESG has developed as a central component of a pilot
assistance system for a military helicopter. The sensor control
software has been developed in accordance with DO-178B, level D and
was accepted by the German military certification authority WTD~61/ML
in 2014.

\begin{figure}
      \centering
      \begin{tikzpicture}[%
        >=latex,
        node distance=7mm,
        basicblock/.style={
          rectangle,
          minimum size=5mm,
          thin,
          draw=black,
          font=\small,
          fill=white
        },
        smallblock/.style={basicblock},
        largeblock/.style={
          basicblock,
          minimum height=20mm,
        }]
        \node[largeblock,draw,fill=white,text width=1.5cm,text centered]
        (assembly) {Monitored Unit};
        \node[smallblock,anchor=west,draw,text centered] at (assembly.35) (ntc1) {NTC 1}; 
        \node[smallblock,anchor=west,draw,text centered] at (assembly.325) (ntc2) {NTC 2}; 
        \node[smallblock,right=of ntc1,draw,text centered] (adc1) {ADC 1};
        \node[smallblock,right=of ntc2,draw,text centered] (adc2) {ADC 2};
        \node[largeblock,text centered] (proc) at (5.5,0) {Processor};
        
        \draw (ntc1) -- (adc1);
        \path[draw] (adc1) -- (proc.west |- adc1);
        \draw (ntc2) -- (adc2);
        \path[draw] (adc2) -- (proc.west |- adc2);

        \begin{pgfonlayer}{background}
          \node[fit=(adc1) (adc2) (proc),thick,draw,fill=black!20,
                inner sep=7pt,label=below:{\small Micro Controller}]{};
        \end{pgfonlayer}
      \end{tikzpicture}
   \caption[]{Temperature Monitoring Hardware}
  \label{fig:temp-mon-hw}
\end{figure}
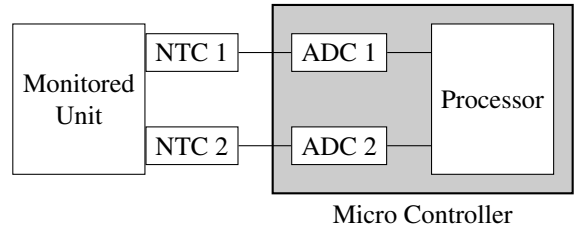

Monitoring the environmental conditions, especially the temperature,
is important to the correct operation of the sensor. Therefore, two
temperature sensors (NTC1 and NTC2 in figure \ref{fig:temp-mon-hw})
are installed within the equipment for redundant temperature
measurements.

The temperature monitoring function is part of the sensor's control
program that is running on the processor shown in figure
\ref{fig:temp-mon-hw}. It calculates the mean temperature of
both readings, as long as they do not differ more than a certain
amount (declared as positive integer \texttt{TEMP\_FAIL}). It accepts
\texttt{MAX\_TEMP\_ERR\_CNT} consecutive differences larger than
\texttt{TEMP\_FAIL} before returning the error \texttt{EC\_TEMP}. 

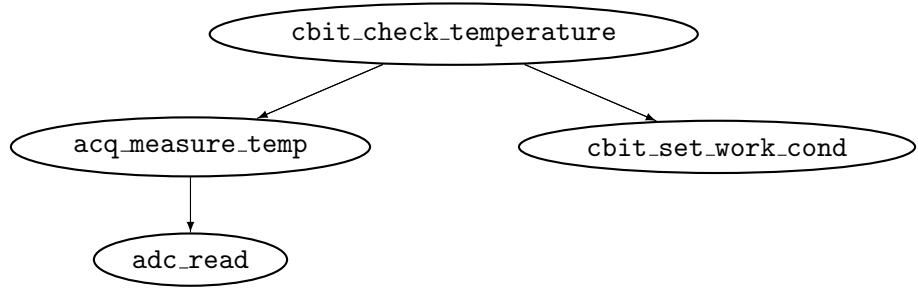
\begin{figure}
      \centering
      \begin{tikzpicture}[%
        >=latex,
        function/.style={
          ellipse,
          minimum size=6mm,
          thick,
          draw=black,
          font=\ttfamily,
        },
        level distance=15mm,
        sibling distance=70mm]
        \node (checktemp) [function,text centered] {cbit\_check\_temperature}
           child{ node (measure) [function,text centered] {acq\_measure\_temp}
              child {node (read) [function,text centered] {adc\_read}} }
           child{node (setworkcond) [function,text centered] {cbit\_set\_work\_cond}
           };
        \path (checktemp) edge[->] (measure);
        \path (checktemp) edge[->] (setworkcond);
        \path (measure) edge[->] (read);
      \end{tikzpicture}
  \caption{Temperature Monitoring Call Hierarchy}
  \label{fig:temp-mon-call-hierarchy}
\end{figure}

Figure \ref{fig:temp-mon-call-hierarchy} shows the temperature
monitoring call hierarchy: the function
\texttt{cbit\_check\_temperature} is executed every
\SI{500}{\milli\second}. This function calls
\texttt{acq\_measure\_temp} that returns the temperature readings of
the two temperature sensors. To do so, the latter function calls
\texttt{adc\_read} to obtain the voltage levels from the
Analog-Digital converters (ADCs, see Fig. \ref{fig:temp-mon-hw})
connected to the temperature sensors.

If the temperature reading is outside operational limits, this fact is
reported as a \emph{working condition} in a module state variable that
is set by the function \texttt{cbit\_set\_working\_cond}.

This  set of functions has been selected for the following reasons:

\begin{itemize}
\item The functions are rather simple but already required some of the
  major concepts of ACSL for their formalization.
\item The selected functions have been taken from a real project, it
  is real code and real specifications with all the flaws that
  typically occur in a time constrained industrial project.
\item The functions cover a standard problem (monitoring a certain
  value obtained from an analogue-to-digital (AD) converter) so it
  does not reveal intellectual property. They are well suited for public
  discussion.
\item The example could well be isolated: the source files contain
  only the selected C functions without failing compilation and
  analysis.
\item The whole set of functions encompasses purely algorithmic as
  well as hardware related functions.
\end{itemize}

\section{Formalization and Verification of LLR}
\label{sec:form-verif-llr}

The analysis approach is the same for all functions: at first, the
original specification of the C function is formalized using ACSL
behavior specifications. In a subsequent step, verification of the
source code is attempted to show by proof that the implementation
follows the specification. This is actually an iterative approach
which often needs adding instrumentation (assertions, loop invariants)
to the source code.

To illustrate the approach, we use the most simple function
\texttt{cbit\_set\_work\_cond} that simply writes to a module
variable, i.e. a persistent variable only known within the module
\texttt{cbit}. This variable is used in other parts of the program but
is read and write protected by setter and getter functions.

The update of the state variable is guarded -- the function takes a
new value of the working condition as a parameter \texttt{new\_cond}
and writes that to the module variable \texttt{WorkCondition} unless
this module variable has been set previously. The only exception is
the special parameter value \texttt{NCD\_NO\_COMMAND} that can
overwrite all settings. A natural language formulation of the low
level requirements would read as follows:

\begin{enumerate}
\item If the value of the input parameter \texttt{new\_cond} is equal
  to \texttt{NCD\_NO\_COMMAND}, then the module variable
  \texttt{WorkCondition} is set to this value
  \texttt{NCD\_NO\_COMMAND}.
    
\item If the current content of the module variable
  \texttt{WorkCondition} is equal to \texttt{NCD\_IDLE\_EMIT}, the
  module variable \texttt{WorkCondition} is set to the value of the
  input parameter \texttt{new\_cond}.

\item If the value of the module variable \texttt{WorkCondition} is
  not equal to \texttt{NCD\_IDLE\_EMIT} and the value of the input
  parameter \texttt{new\_cond} is not equal to
  \texttt{NCD\_NO\_COMMAND}, then the current value of
  \texttt{WorkCondition} is not changed.
\end{enumerate}

The attempt to formalize these LLR into ACSL encounters the first
obstacle: the internal module variable -- a static variable of
file scope -- is not visible in the header
file where the formalized specification is located. The solution was
to introduce a ghost variable that represents the internal data, which
in effect is uncovering the internal variable, i.e. turning parts of
the module inside out. The ghost variable for the internal state is
located in the header file:

\begin{lstlisting}[style=acsl-block]
  //@ ghost uint16_t gWorkCond = NCD_IDLE_EMIT; 
\end{lstlisting}

The behavioral specification, i.e. the translation of the LLR
above, is now straight forward:

\begin{lstlisting}[style=acsl-block]
/*@ 
  @ behavior NoCommand:
  @  assumes new_cond == NCD_NO_COMMAND;
  @  ensures gWorkCond == new_cond;
  @ behavior ModifyWC:
  @  assumes gWorkCond == NCD_IDLE_EMIT;
  @  assumes new_cond != NCD_NO_COMMAND;
  @  ensures gWorkCond == new_cond;
  @ behavior KeepWC:
  @  assumes ((gWorkCond != NCD_IDLE_EMIT) && 
  @           (new_cond != NCD_NO_COMMAND));
  @  ensures gWorkCond == \old(gWorkCond);
  @ complete behaviors;
  @ disjoint behaviors;
  @*/
void cbit_set_work_cond(uint16_t new_cond);
\end{lstlisting}

The three behavior specifications above directly correspond to the
natural language requirements so that they could be replaced by the
formalized requirements. The \texttt{behavior} clause can be named as
shown in the example which facilitates traceability to higher level
requirements, an objective that is strongly emphasized in aviation
standards. The \texttt{assumes} clauses represent the conditional part
of the natural language requirements and are evaluated at the
beginning of the execution of the function. Moreover, with
\texttt{complete} and \texttt{disjoint} clauses one can automatically
verify completeness and consistency of the formal specification.

Note that \texttt{assigns} clauses have been omitted from the
specification above although this is discouraged by the ACSL reference
manual (\cite{Baudin2013}, section 2.3.5). However, adding
\texttt{assigns} clauses generates verification conditions that cannot
be proved. This is due to the fact that the function modifies a state
variable of file scope (\texttt{WorkCondition}) which cannot be listed
in the \texttt{assigns} clause because it is not visible there.
Several solutions for this problem are under discussion on the tool's
mailing list at this time of writing.

For the verification, the ghost state variable must be aligned with
the internal module variable using assertions, as shown below:

\begin{lstlisting}[style=acsl-block]
void cbit_set_work_cond(uint16_t new_cond)
{
  //@ assert gWorkCond == WorkCondition;
    if ((WorkCondition == NCD_IDLE_EMIT) || (new_cond == NCD_NO_COMMAND))
    {
        WorkCondition = new_cond;
        //@ ghost gWorkCond = WorkCondition; 
    }
}
\end{lstlisting}

The assertion in line 3 is necessary to inform the prover that the
internal state is always equal to the ghost state. It is not possible
to prove it.

The WP plugin generates a verification condition for every assertion.
The verification conditions are similar to test cases in traditional
verification so that a verification condition that cannot be proved is
identical to a test case that has failed. Therefore, justification is
needed for the assertion in line 3 to fulfill objective FM2 of table
FM.C-7 of DO-333 that demands that formal analysis results are correct
and discrepancies are explained.

Representing internal state with a ghost variable is not optimal as
indicated above. The function \texttt{cbit\_set\_work\_cond} and its
counterpart \texttt{cbit\_get\_work\_cond} would have been better
specified by using algebraic specification techniques in the way it is
shown in \cite{Burghardt2011} for the stack example. The approach
described there transforms the axioms into additional, ACSL annotated
C code. This essentially means that C code is used as low level
requirements which is not acceptable\footnote{EASA is concerned of
  even using pseudocode as LLR \cite{EASACMSWCEH002}}. The inclusion
of algebraic specification techniques into ACSL would probably be the
most elegant solution for this problem.

\section{Obstacles}
\label{sec:obstacles}
The previous section has already discussed one of the obstacles we
encountered when applying a formal notation to the specification of
LLRs, which is the need to address internal state. There were
additional issues which will be described in the following
subsections.

\subsection{Specifying Behavior across multiple Invocations}
\label{sec:multiple-invocations}

With ACSL, one can only specify a single execution of a C function.
However, the result of a function call sometimes depends on prior
executions of this or even other functions. For example, the function
\texttt{cbit\_check\_temperature} tolerates two consecutive
discrepancies in the readings of the two temperature sensors before
indicating an error condition. This is usually implemented with
internal state variables (\texttt{err\_cnt} in our example), which is
a static variable in the definition of the function, not visible to
the outside. This again cannot be addressed in the specification
located in the header file. As a solution, a ghost variable
\texttt{gerrcnt} has been introduced into the specification, but as in
the previous section, the ghost variable must be updated accordingly
using assertions within the body of the function.

This solution is also not optimal because it is
rather close to the implementation and discloses internals of the
implementation.

A better way in this example is to specify the intended behavior with
a state automaton. The Frama-C plugin Aora\"{i} \cite{Stouls2015} can
be used for this purpose. It translates automaton specifications
formulated in YA or LTL into ACSL annotations (and additional
annotated C code) that can subsequently be verified with the WP
plugin. However, one needs to learn an additional specification
language to use this approach. We were not able to accomplish
this in the time available for the study.

During verification attempt we detected an inaccuracy
in the natural language specification concerning the number of
consecutive temperature discrepancies. The implementation is correct,
but the specification text is misleading.

\subsection{Input Values obtained from Calls to Subroutines}
\label{sec:input-from-subroutines}

As shown in Fig. \ref{fig:temp-mon-call-hierarchy}, function
\texttt{cbit\_check\_temperature} calls \texttt{acq\_temp\_measure} to
obtain temperature readings as input, and \texttt{acq\_temp\_measure}
calls \texttt{adc\_read} to get the voltage levels from the ADCs as
input. This pattern of calling subroutines to obtain input was used a
lot in the real world example. Again, the  return values of
subroutines are  not visible at specification level. 

Ghost variables that represent return values of subroutines were
introduced as a work-around. These ghost variables (\texttt{T1} and
\texttt{T2}) must be aligned with the values of the subroutine call:

\begin{lstlisting}[style=acsl-block]
    acq_measure_temp(&temp1, &temp2);
    //@ ghost T1 = temp1;
    //@ ghost T2 = temp2;
\end{lstlisting}%

Although \texttt{T1} and \texttt{T2} act as input, their values are
obtained during execution and nothing is known at pre-state. It is
therefore not possible to express pre-conditions in the form of
\texttt{assumes} clauses with these input variables.

We introduced predicates to replace the
\texttt{assumes} clauses, as shown in the following code snippet:

\begin{lstlisting}[style=acsl-block]
  @ predicate A_TempReadFailTrans
  @    (integer t1, integer t2, integer cnt) = 
  @    (\abs(t1 - t2) > TEMP_FAIL) && ( cnt <= 2);
  @
  @ predicate A_TempReadFailPerm
  @    (integer t1, integer t2, integer cnt) = 
  @    (\abs(t1 - t2) > TEMP_FAIL) && ( cnt > 2);
\end{lstlisting}%

These predicates are now used as replacement for the \texttt{assumes}
clause as shown in the following specification of the behaviour in
case of discrepancies of temperature readings:

\begin{lstlisting}[style=acsl-block]
  @ ...
  @ behavior TempReadFailTrans:
  @  ensures A_TempReadFailTrans(T1,T2,\at(gerrcnt,Pre)) ==> 
  @            (gModuleTemp == \old(gModuleTemp));
  @  ensures A_TempReadFailTrans(T1,T2,\at(gerrcnt,Pre)) ==>  
  @            (gerrcnt == \old(gerrcnt) + 1);
  @  ensures  A_TempReadFailTrans(T1,T2,\at(gerrcnt,Pre)) ==> 
  @             (\result == EC_NO_ERROR);
  @
  @ behavior TempReadFailPerm:
  @  ensures A_TempReadFailPerm(T1,T2,\at(gerrcnt,Pre)) ==> 
  @            (gModuleTemp == 0);
  @  ensures A_TempReadFailPerm(T1,T2,\at(gerrcnt,Pre)) ==> 
  @            (gerrcnt == \old(gerrcnt));
  @  ensures A_TempReadFailPerm(T1,T2,\at(gerrcnt,Pre)) ==> 
  @            (\result == EC_TEMP);
  @
  @ behavior TempOK:
  @  ensures A_TempReadOK(T1,T2) ==> 
  @            (gModuleTemp == temp_average(T1,T2));
  @  ensures A_TempReadOK(T1,T2) ==> (\result == EC_NO_ERROR);
  @  ensures A_TempReadOK(T1,T2) ==> (gerrcnt == 0);
  @  ...
\end{lstlisting}%

As one can see in the example above, we used a naming
convention to indicate that certain predicates replace
\texttt{assumes} clauses. However, this approach is not very
elegant and has additional disadvantages that are discussed in the
next subsection.%

The verification attempt revealed another weakness in the natural
language specification -- it is ambiguous with respect to the
calculated module temperature in case of permanent temperature
discrepancy. Here again is the implementation correct but the
specification text is misleading.

\subsection{Behavior Specifications}
\label{sec:behav-specs}
Without \texttt{assumes} clauses it is no longer possible to use
\texttt{complete} and \texttt{disjoint} clauses, i.e. it is not
possible anymore to use the tool to simply prove completeness and
consistency (i.e. ``disjointedness'') of the specification. As a
work-around, one  can formulate completeness and consistency
properties with  \texttt{ensures} clauses. Completeness is formulated
as follows:

\begin{lstlisting}[style=acsl-block]
  @ ensures A_TempReadFailTrans(T1,T2,\at(gerrcnt,Pre)) ||
  @    A_TempReadFailPerm(T1,T2,\at(gerrcnt,Pre)) ||
  @    A_TempReadOK(T1,T2);
\end{lstlisting}

Consistency is expressed as 
\begin{lstlisting}[style=acsl-block]
  @ ensures ! (( A_TempReadFailTrans(T1,T2,\at(gerrcnt,Pre)) && 
  @     A_TempReadFailPerm(T1,T2,\at(gerrcnt,Pre))) ||
  @    (A_TempReadFailTrans(T1,T2,\at(gerrcnt,Pre)) && 
  @     A_TempReadOK(T1,T2)) ||
  @    (A_TempReadFailPerm(T1,T2,\at(gerrcnt,Pre)) && 
  @     A_TempReadOK(T1,T2)));
\end{lstlisting}

For consistency specifications with many predicates this easily
becomes complicated and nontransparent.

When taking this problem and the discussion of the previous subsection
into account, it is better to avoid input values obtained from
subroutine calls and to add this as a rule to the design standards. In
our example, the calling function of \texttt{cbit\_check\_temperature}
would call \texttt{adc\_read, acq\_measure\_temp},
\texttt{cbit\_check\_temperature}, and \texttt{cbit\_set\_work\_cond}
in a row. This would also
flatten the call hierarchy which has additional advantages as
explained in the next subsection.

There is a subtle flaw in the original natural language specification
of \texttt{cbit\_check\_temperature} as well as in its formal
counterpart. The function sets the module variable
\texttt{WorkCondition} but this is not specified for all cases. It is
implicitly assumed that it is not modified in these cases (which is
correct), but formally an implementation is free to modify the working
condition by any value. Such omissions can become critical if
specifications are part of interface contracts. The method and the
notation itself cannot prevent such specification errors. The problem
is addressed by DO-333 in objective FM.5-8 of table FM.C-7
(Verification Coverage of Software Structure is achieved). This
objective refers to section FM.6.7.1.3 (Completeness of the Set of
Requirements) that states that for all outputs, the required input
conditions must have been specified. The output that must be
considered here is the internal module variable
\texttt{WorkCondition}.

\subsection{Specification Proliferation}
\label{sec:spec-proliferation}

The original description of \texttt{cbit\_check\_temperature} states
that, if the module temperature is outside the range between
\texttt{TEMP\_MIN} and \texttt{TEMP\_MAX}, the module variable
\texttt{WorkCondition} shall be set to \texttt{NCD\_TEMP\_LOW} or
\texttt{NCD\_TEMP\_HIGH} respectively by calling
\texttt{cbit\_set\_work\_cond} (see section \ref{sec:form-verif-llr}).

This was formalized as follows:
\begin{lstlisting}[style=acsl-block]
  @  ensures (A_TempReadOK(T1,T2) && TempTooCold(T1,T2)) ==> 
  @             (gWorkCond == NCD_TEMP_LOW);
\end{lstlisting}

This cannot be proved because \texttt{cbit\_set\_work\_cond}
only
overwrites the working condition if the current value of the working
condition is equal to \texttt{NCD\_IDLE\_EMIT} (see section
\ref{sec:form-verif-llr}). The corrected specification (which can be
proved) is
\begin{lstlisting}[style=acsl-block]
  @  ensures (A_TempReadOK(T1,T2) && TempTooCold(T1,T2) && 
  @           gWorkCond == NCD_IDLE_EMIT) ==> 
  @                (gWorkCond == NCD_TEMP_LOW);
\end{lstlisting}

This specification repeats parts of the specification of 
\texttt{cbit\_set\_work\_cond}. It is an example of
\emph{specification proliferation} where higher level functions
partially repeat and aggregate those of lower level functions.
It leads to redundancy in the specifications which in turn leads to a
higher maintenance effort, and it counteracts to a certain extent the
well established information hiding heuristic.

One option to alleviate the specification proliferation is keeping the
call hierarchy as flat as possible (e.g. by avoiding input values from
subroutine calls). However, this might not be possible for large and
complex programs. Another option is to move all algorithmic complexity
to the lowest levels of the call hierarchy and use formal
specifications only at these levels. The higher level functions should
have simple logic (at best, only call sequences) that can be easily
verified by code review.

Another alternative is to introduce predicates for the expressions
shared in the specifications. A similar approach using ``specification
macros'' has been used in the Verisoft project for recurring
annotations \cite{Baumann2011}. However, this technique cannot be
demonstrated with the small example used in this study.

\subsection{WP Plugin does not support Math Functions}
\label{sec:math-func-not-supported}

The function \texttt{acq\_measure\_temp} converts the digitized
voltage readings of the temperature dependent resistors (NTC1 and NTC2
in Fig. \ref{fig:temp-mon-hw}) into temperature values. In our first
specification attempt we formulated the post-condition on the output
variables \texttt{temp1} and \texttt{temp2} with a logical function
that used the exponential function \texttt{exp} as shown in the code
extract below:
\begin{lstlisting}[style=acsl-block]
...
//@ ghost uint16_t D1, D2;
...
/*@ logic real R(integer T) = 
  @       10000.0 *\exp(3988.0*(1.0/(T+273.15)-1.0/298.15));
  @
  @ logic real U(integer T) = (5.0*R(T))/(R(T)+5360.0);
  @
  @ logic integer D(integer T) = \floor(250.0*U(T)+0.5);
  @*/
...
/*@
  @ ,,,
  @ ensures D(*temp1) >= D1 > D(*temp1 + 1);
  @ ensures D(*temp2) >= D2 > D(*temp2 + 1);
  @ ...
  @*/
void acq_measure_temp(uint16_t* temp1, uint16_t* temp2);
\end{lstlisting}

The logical function $D(T)$ expresses the relation between the
temperature and the digitized voltage reading. Is is defined by the
temperature characteristic of the NTCs and the electrical circuit
design. The ghost variables \texttt{D1} and \texttt{D2} represent the
digitized voltage readings returned by function \texttt{adc\_read} --
we have used here the same approach as in section
\ref{sec:input-from-subroutines}. The \texttt{ensures} clauses use the
logical function to constrain the values of the output variables
\texttt{temp1} and \texttt{temp2}.

We had to learn that the WP plugin in the version used for the study
did not recognize standard math functions. It is possible to extent the
WP plugin to incorporate own definitions of these functions. This
is not a trivial task and very likely out of scope for SMEs. However,
a specification of math functions would be very useful across many
formal specification projects. Section 3.2 of the ACSL manual
\cite{Baudin2013} announces a library for logic specifications of math
functions which would be very helpful for requirements specifications
as intended here.

In a second attempt we defined the logical function $D(T)$ as a
piece-wise linear approximation with ghost arrays containing the
sampling points. This was accepted by the WP plugin, but the
verification attempt failed due to timeouts. We were not able to
explain the timeouts in the time available for the study.

Note that the implementation of the function
\texttt{acq\_measure\_temp} does not use math functions but iterates
through a look-up table of 100 pre-calculated values.

\subsection{Compiler specific Language Extensions}
\label{sec:comp-specifics}

The source code of the real world example uses compiler specific
language extensions for easier access to hardware registers. These
language extensions are obviously not known in standard C and
therefore not in the WP plugin. There are three possible solutions to
this problem:

\begin{enumerate}
\item Ban compiler specifics by coding standard.
\item Define semantics of compiler specifics to make it known to WP
  plugin. This is very laborious and has not been tried.
\item Only use compiler specifics in very small routines at 
  the lowest
  layer and review these manually.
\end{enumerate}

The last option is probably the most practical solution, and should be
enforced by corresponding design and coding rules.

\subsection{Modelling Hardware for Hardware Access Functions}
\label{sec:hw-access-func}

One property of avionics software is that it needs hardware related
programming (section \ref{sec:frama-c}). Specification of
hardware access functions requires modelling of hardware in ACSL. The
hardware interacts with the external world which operates
independently of the software so that the hardware modifies
the content of program variables in a way that is not visible in
program statements.

ACSL offers so called \emph{volatile ghost variables} to model side
effects such as hardware interaction but this concept was not
implemented in the Fluorine version of Frama-C that was used for the
study.

The solution to this obstacle is the same as in section
\ref{sec:comp-specifics}: all hardware access shall be concentrated
into small subroutines which is common programming practice anyway.
These hardware access functions are formally specified like the other
functions, but then reviewed manually for compliance with the LLR
instead of using automated proof.

The micro controller that was used in the project provided up to
sixteen ADC channels onboard. We modelled the set of ADC channels as a
ghost array and used this model to specify and partially verify that
the correct ADC channels were used for temperature monitoring.

\section{Verification Summary}
\label{sec:verification-summary}
Table \ref{tab:verification-status} provides an overview over the
verification attempt of all four functions considered in this study.

\begin{table}
  \centering
  \caption{Verification Status of Temperature Monitoring Functions}
  \begin{tabular}{lrrrr}
\hline\noalign{\smallskip}
  & \multicolumn{4}{c}{Proof Obligations} \\
Function & scheduled & valid & unknown & timed out \\
\noalign{\smallskip}
\hline
\noalign{\smallskip}
\texttt{cbit\_set\_work\_cond} & 6 & 5 & 1 & 0 \\
\texttt{cbit\_check\_temperature} & 22 & 17 & 5 & 0 \\
\texttt{acq\_measure\_temp} & 237 & 221 & 1 & 15 \\
\texttt{adc\_read} & - & - & - & - \\
\hline    
  \end{tabular}
  \label{tab:verification-status}
\end{table}

The WP plugin could not be executed on \texttt{adc\_read} because of
the non-standard C statements in the source, see section
\ref{sec:hw-access-func}. We were able to explain all proof
obligations for which a proof attempt failed except for the 15 proof
obligations of \texttt{acq\_measure\_temp} that timed out. We were not
able to conclude the analysis of the problem due to lack of time. It is
not uncommon to use alternative provers (which is supported in
Frama-C), but this requires additional effort to understand, install,
configure and use these provers.

\section{Results}
\label{sec:results}

This section matches the experience made in conducting this experiment
against the criteria established in section
\ref{sec:industr-fit-purp}.

\paragraph{Familiarity with Notation and Method.}
\label{sec:famil-with-notat-method}
We  had some exposure to formal methods from
past experience with algebraic specifications and with the formal
notation Z ({\cite{Spivey1998}) as well as his participation in the
research project Verisoft XT, but did not have working knowledge of
ACSL and the Frama-C tool suite.

The ACSL notation is quite close to the C programming language syntax,
a couple of additional language constructs must be familiarized.
However, even the simple example that had been chosen for this study
required some more sophisticated concepts such as ghost variables,
logic functions, axioms and labels. As a consequence, the
specifications are not simple to read for the untrained.

One particular difficulty is to find the correct and most efficient
loop invariant, which however is a common problem to formal proof. 

It should be noted that the amount of instrumentation in form of
assertions and loop invariants is in the same order of magnitude as
the size of the source code, or even exceeds it.  This is to
be expected and in line with DO-178 objectives where all source code
must be traceable to requirements.

\paragraph{Complexity of the Formal  Language.}
\label{sec:compl-form-lang}
The language is very close to C syntax and is therefore familiar to C
programmers. However, for more sophisticated specification tasks, the
available ACSL constructs need a much deeper knowledge.  Some
specification tasks require learning of additional languages (e.g. YA
or LTL for Aora\"{i}).

\paragraph{Training.}
\label{sec:training}
We are  only aware of training sessions at
conferences, and of 
courses and project consultancy that had been offered by the
Fraunhofer FOKUS institute. Also the recently founded company
TrustinSoft offers consultancy for Frama-C.

\paragraph{Information Sources.}
\label{sec:information-sources}
A few online resources are available, mostly tutorials and papers. The
most efficient source of information is the Frama-C mailing list,
accessible via the Frama-C Website \cite{FramaCWebsite} where the
members of the Frama-C community provide fast and accurate advice.

\paragraph{Tool Maturity.}
\label{sec:tool-maturity}

We had difficulties installing the tool set on our Linux distribution
(Archlinux 3.14.4-1) and on Cygwin over Microsoft Windows 7. With
assistance of the tools mailing list and the forum of the Linux
distribution we managed to install a command line version of Frama-C
with the WP plugin and the Alt-Ergo proof engine on the Linux system.
Although problems early on in the tool's usage can have detrimental
effects on an organization's acceptance of the tool, this is
considered a minor problem.

The WP plugin should be extended to include all ACSL features and
support of mathematical functions.

\paragraph{User Guidance provided by Tool.}
\label{sec:user-guid-prov-by-tool}
The tool, even the command line version, highlights the goals that
cannot be proved. It is possible to record the verification conditions
that the tool generates. However, these are formulated in an
intermediate language as input to the prover which is quite different
to the C or even ACSL syntax. Analysis of verification conditions
requires deep knowledge of automated prover technologies.

We were not able to prove all generated proof obligations and could
not explain all discrepancies in the time that was allocated for this
study. It may be possible that the Frama-C GUI together with
interactive provers provide more debugging support -- which we could
not test due to the aforementioned installation problems.

\section{Conclusion}
\label{sec:conclusion}
During analysis several smaller deficiencies were discovered in the
low-level requirements. The formalization of natural
language requirements into ACSL is an excellent tool to ``debug''
specifications and to improve their quality.  Moreover, the
formalization helps to achieve many of the DO-178C objectives related
to LLR.

At this time of writing, we recommend to use ACSL and Frama-C
only in highly safety critical applications and even there only in areas
that need extra scrutiny. And even so, the team must be supported by
an experienced consultant for both tool and method who can provide
immediate assistance in case specification or verification problems
occur. We also suggest to collect and publish a set of
specification patterns or best practices similar to those for Z
\cite{Stepney2003,Valentine2004}.

If formal specification is used, then additional rules must be added
to the software design and coding standards in order to facilitate
formal verification. Such rules would include the reduction of the
call hierarchy, the reduction of internal state where possible and the
increase of parameter passing as main method of data flow. 
The decision to use ACSL for specification and verification will shape
the software design and coding practice.

Since formal methods have made their way into aviation standards and
guidelines it is to be expected that they will be mandated in certain
areas in the future, if not by the certification authorities, then by
major aerospace companies like Airbus and Dassault. Since the
WP plugin did not fully implement the ACSL language standard at the
time of the experiment, we also recommend to observe the further
development and to repeat this study in one or two years.

\bibliographystyle{eptcs}
\bibliography{fide2015paper6}

\begin{thebibliography}{10}
\providecommand{\bibitemdeclare}[2]{}
\providecommand{\surnamestart}{}
\providecommand{\surnameend}{}
\providecommand{\urlprefix}{Available at }
\providecommand{\url}[1]{\texttt{#1}}
\providecommand{\href}[2]{\texttt{#2}}
\providecommand{\urlalt}[2]{\href{#1}{#2}}
\providecommand{\doi}[1]{doi:\urlalt{http://dx.doi.org/#1}{#1}}
\providecommand{\bibinfo}[2]{#2}

\bibitemdeclare{techreport}{EASACMSWCEH002}
\bibitem{EASACMSWCEH002}
 (\bibinfo{year}{2012}): \emph{\bibinfo{title}{{EASA Certification Memo
  CM-SWCEH-002 Software Aspects of Certification}}}.
\newblock \bibinfo{type}{Technical Report} \bibinfo{number}{EASA CM-SWCEH002
  Issue 01 Revision 1}, \bibinfo{institution}{European Aviation Safety Agency}.

\bibitemdeclare{misc}{ACSLWeb}
\bibitem{ACSLWeb}
\emph{\bibinfo{title}{{ANSI/ISO} {C Specification Language}}}.
\newblock \urlprefix\url{http://frama-c.com/acsl.html}.

\bibitemdeclare{manual}{Baudin2013a}
\bibitem{Baudin2013a}
\bibinfo{author}{Patrick \surnamestart Baudin\surnameend},
  \bibinfo{author}{Loic \surnamestart Correnson\surnameend} \&
  \bibinfo{author}{Zaynah \surnamestart Dargaye\surnameend}
  (\bibinfo{year}{2013}): \emph{\bibinfo{title}{WP Plug-in Manual. Version 0.7
  for Fluorine-20130601}}.
\newblock \bibinfo{organization}{CEA LIST}.

\bibitemdeclare{techreport}{Baudin2013}
\bibitem{Baudin2013}
\bibinfo{author}{Patrick \surnamestart Baudin\surnameend},
  \bibinfo{author}{Pascal \surnamestart Cuoq\surnameend},
  \bibinfo{author}{Jean-Christophe \surnamestart Filli\^atre\surnameend},
  \bibinfo{author}{Claude \surnamestart March\'e\surnameend},
  \bibinfo{author}{Benjamin \surnamestart Monate\surnameend},
  \bibinfo{author}{Yannick \surnamestart Moy\surnameend} \&
  \bibinfo{author}{Virgile \surnamestart Prevosto\surnameend}
  (\bibinfo{year}{2013}): \emph{\bibinfo{title}{{ACSL}: {ANSI/ISO} {C}
  {Specification Language}. {Version} 1.7}}.
\newblock \bibinfo{type}{Technical Report}, \bibinfo{institution}{CEA LIST,
  Software Reliability Laboratory}.
\newblock \urlprefix\url{http://frama-c.com/download/acsl.pdf}.

\bibitemdeclare{inproceedings}{Baumann2011}
\bibitem{Baumann2011}
\bibinfo{author}{C.~\surnamestart Baumann\surnameend},
  \bibinfo{author}{T.~\surnamestart Bormer\surnameend},
  \bibinfo{author}{H.~\surnamestart Blasum\surnameend} \&
  \bibinfo{author}{S.~\surnamestart Tverdyshev\surnameend}
  (\bibinfo{year}{2011}): \emph{\bibinfo{title}{Proving Memory Separation in a
  Microkernel by Code Level Verification}}.
\newblock In: {\sl \bibinfo{booktitle}{Object/Component/Service-Oriented
  Real-Time Distributed Computing Workshops (ISORCW), 2011 14th IEEE
  International Symposium on}}, pp. \bibinfo{pages}{25--32},
  \doi{10.1109/ISORCW.2011.14}.

\bibitemdeclare{inproceedings}{Baumann2009}
\bibitem{Baumann2009}
\bibinfo{author}{Christoph \surnamestart Baumann\surnameend},
  \bibinfo{author}{Bernhard \surnamestart Beckert\surnameend},
  \bibinfo{author}{Holger \surnamestart Blasum\surnameend} \&
  \bibinfo{author}{Thorsten \surnamestart Bormer\surnameend}
  (\bibinfo{year}{2009}): \emph{\bibinfo{title}{Better Avionics Software
  Reliability by Code Verification -- {A} Glance at Code Verification
  Methodology in the {Verisoft~XT} Project}}.
\newblock In: {\sl \bibinfo{booktitle}{Embedded World 2009 Conference}},
  \bibinfo{publisher}{Franzis Verlag}, \bibinfo{address}{Nuremberg, Germany}.

\bibitemdeclare{inproceedings}{Blasum2012}
\bibitem{Blasum2012}
\bibinfo{author}{Holger \surnamestart Blasum\surnameend},
  \bibinfo{author}{Frank \surnamestart Dordowsky\surnameend},
  \bibinfo{author}{Bruno \surnamestart Langenstein\surnameend} \&
  \bibinfo{author}{Andreas \surnamestart Nonnengart\surnameend}
  (\bibinfo{year}{2012}): \emph{\bibinfo{title}{{DO-178C} Compliance of
  {Verisoft} Formal Methods}}.
\newblock In: {\sl \bibinfo{booktitle}{Proceedings of the Embedded Real Time
  Software and Systems Conference, 1. - 3. February, Toulouse}}.

\bibitemdeclare{techreport}{Burghardt2011}
\bibitem{Burghardt2011}
\bibinfo{author}{Jochen \surnamestart Burghardt\surnameend},
  \bibinfo{author}{Jens~Gerlach \surnamestart amd Liangliang~Gu\surnameend},
  \bibinfo{author}{Kerstin \surnamestart Hartig\surnameend},
  \bibinfo{author}{Hans \surnamestart Pohl\surnameend}, \bibinfo{author}{Juan
  \surnamestart Soto\surnameend} \& \bibinfo{author}{Kim \surnamestart
  VÂ¨ollinger\surnameend} (\bibinfo{year}{2011}): \emph{\bibinfo{title}{{ACSL
  By Example}. {Towards} a Verified {C} Standard Library}}.
\newblock \bibinfo{type}{Technical Report}, \bibinfo{institution}{Fraunhofer
  FIRST}.

\bibitemdeclare{}{CCPart3}
\bibitem{CCPart3}
 (\bibinfo{year}{2007}): \emph{\bibinfo{title}{{Common Criteria for Information
  Technology Security Evaluation. Part 3: Security Assurance Components}}}.

\bibitemdeclare{manual}{Correnson2013}
\bibitem{Correnson2013}
\bibinfo{author}{Lo\"{i}c \surnamestart Correnson\surnameend},
  \bibinfo{author}{Pascal \surnamestart Cuoq\surnameend},
  \bibinfo{author}{Florent \surnamestart Kirchner\surnameend},
  \bibinfo{author}{Virgile \surnamestart Prevosto\surnameend},
  \bibinfo{author}{Armand \surnamestart Puccetti\surnameend},
  \bibinfo{author}{Julien \surnamestart Signoles\surnameend} \&
  \bibinfo{author}{Boris \surnamestart Yakobowski\surnameend}
  (\bibinfo{year}{2013}): \emph{\bibinfo{title}{Frama-C User Manual. Release
  Fluorine-20130601}}.
\newblock \urlprefix\url{http://frama-c.com/download/frama-c-user-manual.pdf}.

\bibitemdeclare{}{DO178C}
\bibitem{DO178C}
 (\bibinfo{year}{2011}): \emph{\bibinfo{title}{{RTCA DO-178C} {Software
  Considerations in Airborne Systems and Equipment Certification}}}.

\bibitemdeclare{}{DO333}
\bibitem{DO333}
 (\bibinfo{year}{2011}): \emph{\bibinfo{title}{{RTCA DO-333} {Formal Methods
  Supplement to DO-178C and DO-278A}}}.

\bibitemdeclare{misc}{FramaCWebsite}
\bibitem{FramaCWebsite}
\emph{\bibinfo{title}{{Frama-C} Software Analyzers}}.
\newblock \urlprefix\url{http://frama-c.com/}.

\bibitemdeclare{techreport}{Rushby1993}
\bibitem{Rushby1993}
\bibinfo{author}{John \surnamestart Rushby\surnameend} (\bibinfo{year}{1993}):
  \emph{\bibinfo{title}{Formal Methods and the Certification of Critical
  Systems}}.
\newblock \bibinfo{type}{Technical Report} \bibinfo{number}{{SRI-CSL-93-7}},
  \bibinfo{institution}{Computer Science Laboratory, {SRI} International},
  \bibinfo{address}{Menlo Park, {CA}}.

\bibitemdeclare{incollection}{Souyris2009}
\bibitem{Souyris2009}
\bibinfo{author}{Jean \surnamestart Souyris\surnameend},
  \bibinfo{author}{Virginie \surnamestart Wiels\surnameend},
  \bibinfo{author}{David \surnamestart Delmas\surnameend} \&
  \bibinfo{author}{Herv{\'{e}} \surnamestart Delseny\surnameend}
  (\bibinfo{year}{2009}): \emph{\bibinfo{title}{Formal Verification of Avionics
  Software Products}}.
\newblock In \bibinfo{editor}{Ana \surnamestart Cavalcanti\surnameend} \&
  \bibinfo{editor}{Dennis \surnamestart Dams\surnameend}, editors: {\sl
  \bibinfo{booktitle}{FM 2009: Formal Methods}}, {\sl \bibinfo{series}{Lecture
  Notes in Computer Science}} \bibinfo{volume}{5850},
  \bibinfo{publisher}{Springer Berlin / Heidelberg}, pp.
  \bibinfo{pages}{532--546}, \doi{10.1007/978-3-642-05089-3_34}.

\bibitemdeclare{book}{Spivey1998}
\bibitem{Spivey1998}
\bibinfo{author}{J.~M. \surnamestart Spivey\surnameend} (\bibinfo{year}{1998}):
  \emph{\bibinfo{title}{The Z Notation: A Reference Manual}},
  \bibinfo{edition}{2nd edition} edition.
\newblock \bibinfo{publisher}{Prentice Hall International (UK) Ltd}.

\bibitemdeclare{techreport}{Stepney2003}
\bibitem{Stepney2003}
\bibinfo{author}{Susan \surnamestart Stepney\surnameend},
  \bibinfo{author}{Fiona \surnamestart Polack\surnameend} \&
  \bibinfo{author}{Ian \surnamestart Toyn\surnameend} (\bibinfo{year}{2003}):
  \emph{\bibinfo{title}{A {Z} Patterns Catalogue {I}: Specification and
  Refactorings}}.
\newblock \bibinfo{type}{Technical Report} \bibinfo{number}{YCS-2003-349},
  \bibinfo{institution}{Department of Computer Science, University of York}.

\bibitemdeclare{techreport}{Stouls2015}
\bibitem{Stouls2015}
\bibinfo{author}{Nicolas \surnamestart Stouls\surnameend} \&
  \bibinfo{author}{Virgile \surnamestart Prevosto\surnameend}
  (\bibinfo{year}{2015}): \emph{\bibinfo{title}{Aorai Plugin Tutorial}}.
\newblock \bibinfo{type}{Technical Report}, \bibinfo{institution}{INRIA}.
\newblock \urlprefix\url{http://frama-c.com/download/frama-c-aorai-manual.pdf}.

\bibitemdeclare{techreport}{Valentine2004}
\bibitem{Valentine2004}
\bibinfo{author}{Samuel~H. \surnamestart Valentine\surnameend},
  \bibinfo{author}{Susan \surnamestart Stepney\surnameend} \&
  \bibinfo{author}{Ian \surnamestart Toyn\surnameend} (\bibinfo{year}{2004}):
  \emph{\bibinfo{title}{A {Z} Patterns Catalogue {II}: Definitions and Laws}}.
\newblock \bibinfo{type}{Technical Report} \bibinfo{number}{YCS-2004-383},
  \bibinfo{institution}{Department of Computer Science, University of York}.

\bibitemdeclare{misc}{VCCWebsite}
\bibitem{VCCWebsite}
\emph{\bibinfo{title}{{VCC} Website}}.
\newblock \urlprefix\url{http://vcc.codeplex.com/}.

\bibitemdeclare{techreport}{Vries1996}
\bibitem{Vries1996}
\bibinfo{author}{L.M.G. \surnamestart de~Vries\surnameend}
  (\bibinfo{year}{1996}): \emph{\bibinfo{title}{Applying Formal Methods in the
  {DO-178B} Certification Process}}.
\newblock \bibinfo{type}{Technical Report} \bibinfo{number}{NLR TP 95547},
  \bibinfo{institution}{National Aerospace Laboratory NLR, Amsterdam, The
  Netherlands}.

\end{thebibliography}

\end{document}